# REMEMBERING GUIDO ALTARELLI (*)

Mario Greco

Department of Mathematics and Physics and INFN, University of Roma Tre

I have been asked to organize this session in memory of Guido Altarelli, who participated to the last edition of this astro-particle physics conference in 2014, and I have done it with great pleasure and sadness at the same time, because of my more than fifty-years-long friendship with Guido. We first met when students at the University of Rome in the early 1960's, and in the following many more years we got quite close friends together with our families. Often we discussed physical problems of common interest but we also shared together some nice vacation times. Our offices at the Physics Department of Roma Tre University have been just a few meters away.

In this talk I will recall the early times of his scientific career and his major role played in the development of QCD, in particular in those fields I have been also involved, leaving to the following speakers to describe other aspects of his research activity and also the role he played in the development of phenomenology of particle physics, particularly at CERN. Guido himself has written his own recollections on the early days of QCD in Rome in the 70's and early 80's, in the occasion of my $70^{th}$ anniversary [1]. A similar description of the contributions of Guido in the evolution of QCD has been presented by Keith Ellis last March at the 2016 La Thuile conference [2].

Together with Guido Altarelli, I recall many other friends and future colleagues at the university times, as Franco Buccella, Giorgio Capon, Sergio Doplicher, Giovanni Gallavotti, Adalberto Giazotto, Luciano Maiani, Francesco Melchiorri, Piergiorgio Picozza, and Piero Spillantini. For sure many of them are well known to this audience of astro-particle physicists.

Many people know that Guido worked for his diploma thesis together with Franco Buccella, whom he knew since the primary schools, on the problem of single bremsstrahlung emission in electron-positron annihilation [3], suggested to them by Raoul Gatto. Indeed in those years, after Bruno Touschek's seminal idea of colliding beams, the construction at Frascati of AdA (in 1961, with an energy of 250 MeV) and later of ADONE (with an energy of 3 GeV) led much theoretical efforts in the roman area to investigate the physics of electron-positron annihilation. Actually Guido did his first attempt for his diploma thesis with Bruno Touschek, who was at the time very busy with the luminosity problems of AdA which, as it is well known, was finally moved to Orsay to improve the injection. In a cold day of autumn Guido finally succeeded in getting an appointment with him in his office, but because of a malfunctioning heater, Touschek's jacket started burning, he got very nervous and the talk ended poorly.

After his degree, Guido went to Florence joining the new Gatto's theoretical group, with other roman colleagues and friends, as Franco Buccella, Giovanni Gallavotti, Luciano Maiani and Giuliano Preparata, forming the group of the so called "gattini" (meaning kittens). They had been working in those years mainly on unitary symmetries and super-convergence relations in strong interactions. On the other hand, with the advent of ADONE, a sizeable theory group developed at Frascati, whose members were Gianni De Franceschi, Paolo Di Vecchia, Etim Etim, Giulia Pancheri, Giancarlo Rossi and myself, under the direction of Bruno Touschek, who was quite convinced that a proper quantitative analysis of the experimental results from an $e^+e^-$ colliding beam needed a precise computation of QED radiative corrections re-summed to all orders. Some of the basic ideas and techniques developed there will be of great help in later years in the understanding and the description of resummation effects in QCD.

In 1968-70 Guido spent two academic years at NYU and Rockfeller University working on S-Matrix duality [4] and light-cone expansions [5] with R. Brandt and G. Preparata. The general physics context at that time was the approximate scaling properties shown by the SLAC deep inelastic scattering data, which had motivated the Feynman's parton model [6] as well as the study of the properties of the commutator of the electromagnetic currents on the light cone [7]. On the other hand Leon Lederman's group had presented their results [8] on the production of muon pairs in hadronic collisions and immediately later S. Drell and T. Yan [9] had proposed the quark-antiquark annihilation mechanism. At Frascati in the meantime ADONE was producing the first tests of QED in the GeV region and the early results on the large multi-hadron production [10].

Once back to Rome Guido started working with Nicola Cabibbo and Luciano Maiani on various subjects: sigma terms, chiral symmetry and scale invariance [11], deep inelastic phenomena [12], and after the discovery of the asymptotic freedom in gauge theories [13,14], on the octet enhancement of non-leptonic weak interactions [15].

The discovery of the J/ψ [16,17] in November 1974 was a shock in Rome as elsewhere. At Frascati the ADONE accelerator team was able to raise the energy up to 3.1 GeV to produce the J/ψ a couple of days after SLAC, and publish their results in the same PRL issue [18]. Unfortunately, a private communication with the experimenters indicating an apparent forward-backward asymmetry observed in the muon pairs produced on resonance, gave rise to a paper from the Rome group [19], who wrongly concluded that the particle discovered was the $Z_0$ boson. On the other hand in that period of time I had the terrific chance to be invited by Sid Drell to give a seminar on my previous works on duality, and arrived at SLAC just the day after the discovery. Those were days of great excitement, endless talks, and after some detailed discussions on the data with Burt Richter and Roy Schwitters a few days later I left to Mexico City which was my final destination. The careful study of the J/ψ data and the subsequent discovery of the ψ', which I learnt from a local newspaper, immediately led Cesareo Dominguez and myself, using duality ideas, to interpret the new particles as a new series of bound states of charm quark-antiquark pairs and also predict the appropriate increase of the famous ratio R after the open charm threshold

[20], as independently suggested at the same time by Appelquist and Politzer [21] and De Rujula and Glashow [22].

Let's discuss now the great progress in the development of QCD realized with Altarelli – Parisi (AP) equations [23]. Before that, the applications of QCD to physical processes was quite complicated at least for two reasons. By using Guido's words, *"QCD is a theory of quarks and gluons while only hadrons are observable. Moreover perturbation theory can only be applied in those particular domains of the strong interaction where approximate freedom, which is only asymptotic, can be reached"* [23]. In addition, *"in spite of the relative simplicity of the final results, their derivation, although theoretically rigorous, is somewhat abstract and formal, being formulated in the language of renormalisation group equations for the coefficient functions of the local operators which appear in the light cone expansion for the product of two currents."*

In 1976 Guido Altarelli and Giorgio Parisi were both on sabbatical leave in Paris, Guido at the Ecole Normale Superieure and Giorgio at the Institut des Hautes Etudes Scientifiques at Bures-sur-Yvette. Giorgio had presented at the Moriond conference a paper [24] containing an early form of the AP equation, namely a simple evolution equation for the electron and photon distributions after a bremsstrahlung emission, suggesting a similar treatment in QCD. The common paper appeared in 1977 [23] and the main virtue of their approach was to formulate the evolution of the parton densities as a branching process with probabilities determined (at leading order) by the splitting functions (proportional to the running coupling). A particular emphasis was devoted to prove that the splitting functions are a property of the theory and do not depend on the process, in particular the evolution does not apply only to deep inelastic scattering.

Within the general framework given by the AP equations, the Rome group contributed very much to the theory of Drell-Yan processes. In particular an important progress was made in '78-'79 with the calculation of the next to the leading (NLO) corrections to Drell-Yan processes by Guido with Keith Ellis and Guido Martinelli [25]. This was one of the first calculations of NLO corrections in QCD. They started by defining the quark parton densities beyond leading order in a precise way (for quarks they adopted the structure function $F_2$ as the defining quantity, gluons only enter at NLO in Drell-Yan processes). Then the calculation of NLO diagrams for both deep inelastic scattering and the Drell-Yan process allowed to derive the corrective terms for the Drell-Yan cross-section, as function of $Q^2$. The resulting corrections turned out to be surprisingly large. The ratio of corrected to uncorrected (Born) cross-sections was found to be rather constant in $Q^2$ and in rapidity. They denoted it as the "K-factor". This large correction was clearly casting doubts on the convergence of perturbation theory.

Actually the origin of the main part of this correction could be traced back to effects that can be re-summed to all orders. Indeed in a series of works, the tools developed over the years for QED resummations at Frascati, were applied by us to the resummation of soft gluons in different QCD processes, in particular in collaboration with Giuseppe Curci and Yogi Srivastava [26,27]. This concerns the so called "large $\pi^2$ terms" and also the resummations near the phase space boundaries, both in deep inelastic scattering near $x = 1$ and in the Drell-Yan processes near $\tau = Q^2/s = 1$ [28]. Giorgio Parisi also studied the $\pi^2$ problem at near the same time [29].

In Fig.1 it is shown how we looked like in those years.

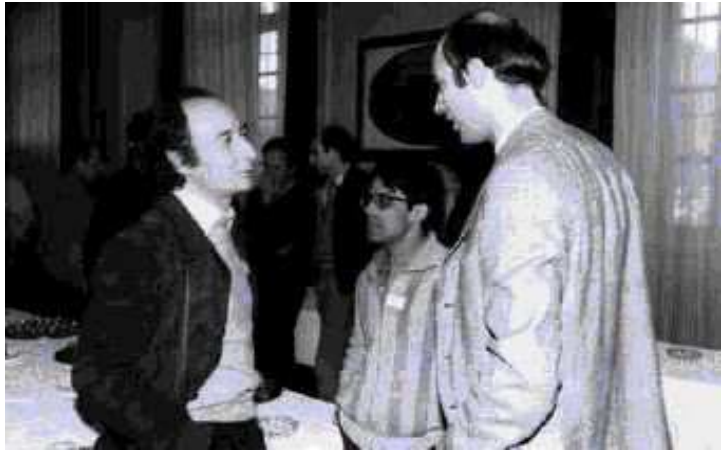

Fig. 1. – From the left: Mario Greco, Yogi Srivastava and Guido Altarelli in 1979 at the Accademia dei Lincei, Rome

Another important theoretical problem for Drell-Yan processes that was attacked in those years is the evaluation of the transverse momentum ($p_T$) distribution of the produced virtual boson (a $\gamma$ or a $W^\pm$ or a $Z_0$). After the LO perturbative result, valid for $p_T \sim Q$, completed in '78 by Guido with Parisi and Petronzio [30], the NLO perturbative calculation was obtained in '81-'83 by K. Ellis, Martinelli and Petronzio [31]. That followed the study of the resummation of the Sudakov double logarithms, paper by Dokshitser, Dyakonov and Troyan [32] who however obtained an incorrect result. Then the correct re-summed answer was given in '79 by Curci, Srivastava and myself [27] taking into account the conservation of the transverse momentum in the multi-gluon emission in the initial state. As soon as the data on the W and Z production from UA1 and UA2 at CERN proton-antiproton collider were available, an adequate theoretical prediction, adding the result of the Sudakov resummation to the complete one loop calculation was finally obtained in a paper by Guido, K. Ellis, G. Martinelli and myself [33]. Using Guido's words [1] "*this is an important paper, because it essentially contained all the crucial ingredients that describe the physics of this phenomenon. In the subsequent years the accuracy was much improved with the computation of sub-leading effects and with several different refinements, but the essential points were all present in our paper and the accuracy of our treatment was adequate for the quality of the first data. The same techniques are at present applied to the calculation of the $p_T$ distribution of the Higgs boson produced by gluon fusion (see, for example, ref. [34])*".

In Fig. 2 all four of us are shown, participating into a conference in Crete in 1980, with thanks to Guido Martinelli.

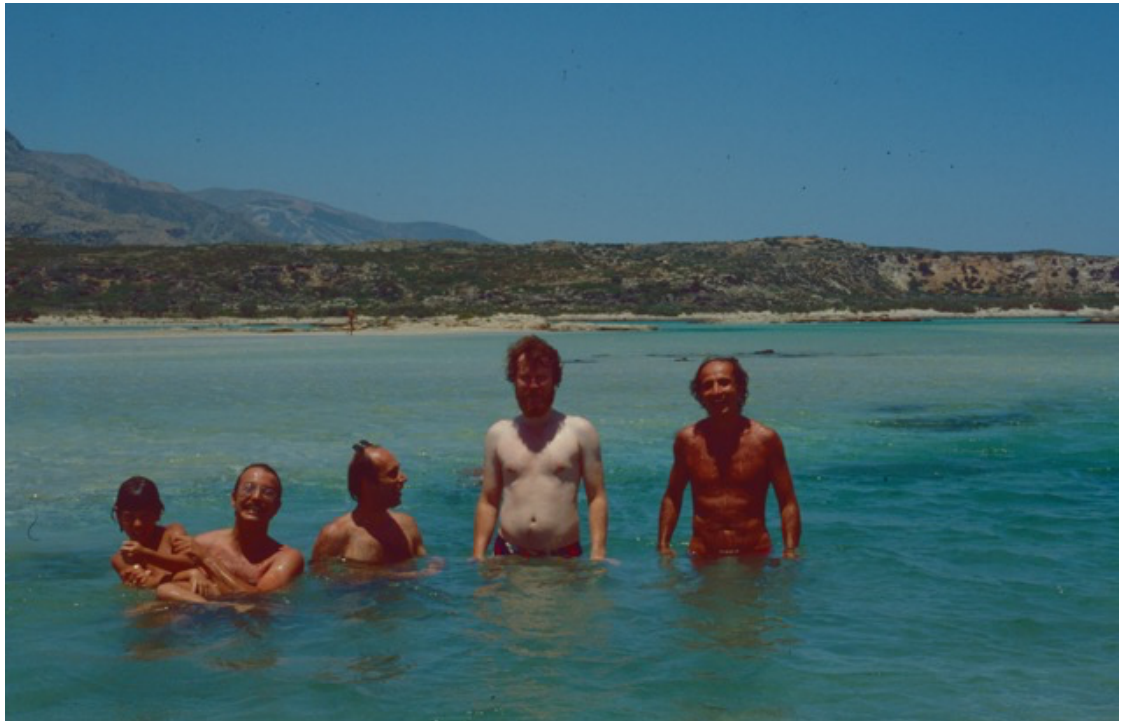

Fig.2 – From the left: Guido Martinelli, Guido Altarelli, Keith Ellis and Mario Greco in Chania, Crete, 1980.

I will stop here, not mentioning further research activity in QCD as, for example, in the small-x physics domain, where both of us have been involved in recent years [35,36], leaving the floor to next speakers. As I already mentioned in the beginning, Guido was so kind to give a talk at La Thuile conference in 2011 on the occasion of my 70$^{th}$ birthday. Fortunately that talk had been recorded and, to conclude, I will show now a few pieces of his presentation in order to leave with us a last image of his impressive personality and kindness.

REFERENCES.